\PassOptionsToPackage{
  a4paper,
  top=18mm,
  bottom=28mm,
  left=18mm,
  right=18mm
}{geometry}
\documentclass[pdflatex,sn-mathphys-num,iicol]{sn-jnl}

\usepackage{graphicx}%
\usepackage{multirow}%
\usepackage{amsmath,amssymb,amsfonts}%
\usepackage{amsthm}%
\usepackage{mathrsfs}%
\usepackage[title]{appendix}%
\usepackage{xcolor}%
\usepackage{textcomp}%
\usepackage{manyfoot}%
\usepackage{booktabs}%
\usepackage{algorithm}%
\usepackage{algorithmicx}%
\usepackage{algpseudocode}%
\usepackage{listings}%

\theoremstyle{thmstyleone}%
\theoremstyle{thmstyletwo}%

\theoremstyle{thmstylethree}%

\begin{document}

\title[Higgs physics at a $\sqrt{s} = 10$ TeV Muon Collider]{\textbf{Higgs physics at a $\mathbf{\sqrt{s} = 10}$  TeV Muon Collider}}
\subtitle{\small Presented at the 32nd International Symposium on Lepton Photon Interactions at High Energies \newline Madison, Wisconsin, USA, August 25-29, 2025}

\author[a]{Paolo Andreetto}
\author[b]{Massimo Casarsa}
\author[a]{Alessio Gianelle}
\author[b,c]{Carlo Giraldin}
\author[a,d]{Donatella Lucchesi}
\author[a,d]{Leonardo Palombini}
\author[e]{Lorenzo Sestini}
\author[a,d]{Davide Zuliani}

\affil[]{\textit{On behalf of the International Muon Collider Collaboration}}

\affil[]{}
\affil[a]{\small INFN Sezione di Padova, Padua, Italy}
\affil[b]{\small INFN Sezione di Trieste, Trieste, Italy}
\affil[c]{\small Universit\`a di Trieste, Trieste, Italy}
\affil[d]{\small Universit\`a di Padova, Padua, Italy}
\affil[e]{\small INFN Sezione di Firenze, Florence, Italy}
\affil[]{\newline  \small Contacts: Leonardo Palombini (\href{leonardo.palombini@pd.infn.it}{leonardo.palombini@pd.infn.it})
}


\abstract{\unboldmath This contribution discusses the physics potential of a future muon collider operating at a center-of-mass energy of $\sqrt{s} = 10$ TeV for precision studies in the Higgs sector. Using a detailed detector simulation that incorporates the dominant sources of machine-induced background, the expected sensitivity to key Higgs processes is evaluated. These include the measurement of production cross sections for 
$H\to bb$, $H\to WW^*$, and double-Higgs production $H\!H\to b\overline{b}b\overline{b}$. A central focus of the study is the determination of the Higgs boson trilinear self-coupling, a critical parameter for understanding the structure of the Higgs potential and electroweak symmetry breaking. The analysis is based on the MUSIC multi-purpose detector concept, specifically optimized for the muon collider environment, and assumes an integrated luminosity of $10$ ab$^{-1}$ collected over five years. The results presented highlight the exceptional prospects of a multi-TeV muon collider for exploring the Higgs potential with a level of precision unattainable by any other proposed future collider within a comparable timeframe.}

\maketitle

\section{Precision Higgs physics at a $\mathbf{\sqrt s = 10}$ TeV Muon Collider}

\noindent
The 2012 discovery of the Higgs boson by the CMS and ATLAS collaborations \cite{atlasHig}\cite{cmsHig} completed the picture of the Standard Model of particle physics (SM), and brought the field of Higgs physics to its \textit{precision era}. A complete characterization of the Higgs sector could produce evidence for new phenomena beyond the Standard Model: 
this includes precise measurements of the couplings of the Higgs boson to the other SM particles and the direct determination of the trilinear Higgs self-coupling.

\noindent
With the High Luminosity phase of the LHC, ATLAS and CMS are set to measure the main Higgs-SM couplings down to the $\%$ level and to constrain the Higgs trilinear coupling with an uncertainty of $\sim 30\%$ \cite{cmsatlas}. However, as highlighted in the 2020 European Strategy for Particle Physics Update \cite{esppu}, a substantial improvement in the knowledge of the Higgs sector should be the priority for the next generation of collider machines. In this context, a future Muon Collider stands out as one of the strongest options.

\noindent
A Muon Collider (MuC) will accelerate and collide beams of muons and anti-muons at multi-TeV center of mass energies. For $\sqrt s = 10$ TeV, considered in this study, a MuC is expected to produce a sample of $\sim 10^7$ single-Higgs events and $\sim 4 \cdot10^4$ double-Higgs events in a relatively clean environment, with an integrated luminosity $\mathcal{L} = 10$ ab$^{-1}$, which can be collected by one experiment in 5 years of operation \cite{towardsmuc}. 

\noindent
The possibility of reaching high sensitivities in the Higgs sector relies on the ability to keep the effects of the machine-induced background under control. 
A detector concept, MUSIC, was designed and optimized for the collision environment of a $\sqrt s = 10$ TeV MuC \cite{music-det-esppu}. The MUSIC concept is used to perform detailed simulation studies, with the aim of evaluating the reachable sensitivities on selected Higgs observables. All simulations include the effects of the dominant machine-induced backgrounds. This contribution discusses the attainable statistical sensitivities on the cross-section times branching ratio ($\sigma\cdot BR$) of the processes $H \rightarrow b \overline b$, $H \rightarrow WW^*$, and $HH \rightarrow b \overline b b \overline b$, and on the measurement of the Higgs trilinear coupling modifier $\kappa_3 = \lambda_3 / \lambda_3^{SM}$. All quoted results consider the situation of 2 interaction points, in which each experiment collects an integrated luminosity of $10$ ab$^{-1}$.

\section{Methodology}
\noindent
A MuC presents unique experimental challenges, mainly related to the in-flight decay of muons in the machine. The machine-induced backgrounds and their impact on detector performance for a $\sqrt s = 10$ TeV MuC have been studied in Ref. \cite{music-det-esppu, BIBIP-calzolari}. The results in this contribution use a detailed detector simulation, based on \textsc{Geant4}~\cite{ref:geant4}, including machine-induced backgrounds generated with the FLUKA software~\cite{fluka}, as described in Ref.~\cite{Lucchesi:2024z+}.

\noindent
The Monte Carlo samples for the signal processes and their associated physics backgrounds are produced with the WHIZARD generator~\cite{whizard}, and the particle hadronization and showering are performed with PYTHIA version 8.200~\cite{pythia}.
Minimal requirements are applied at the generation level to guarantee the absence of biases in the measurements.

\noindent
The samples are successively processed with the simulation and reconstruction tools of the Muon Collider software framework~\cite{MuCSoftware}. 
The simulated energy depositions produced by the machine-induced backgrounds are superimposed on each of the physics processes, event by event, before the event is reconstructed. The MUSIC detector, as well as the reconstruction and identification algorithms for key physics objects and their performance, are described in detail in Ref.~\cite{music-det-esppu}.

\section{Analysis and results}
\noindent
The Monte Carlo samples for each considered final state are analyzed to obtain the statistical sensitivity $\Delta (\sigma\cdot BR)_{stat} / (\sigma \cdot BR)$. The sensitivity on the trilinear coupling modifier $\kappa_3$ is evaluated only in the $HH\rightarrow b \overline b b \overline b$ decay channel. All analyses assume an integrated luminosity $\mathcal{L} = 10$ ab$^{-1}$, equivalent to 5 years of data taking for one experiment.

\begin{table}[t]
    \centering
    \caption{Generation cross section, preselection efficiency, $b$-tagging efficiency, and expected events for the signal and physics background processes in the $H \to b \bar{b}$ analysis for $\mu^+\mu^-$ collisions at 10 TeV and $\mathcal{L}=10$ ab$^{-1}$. $X$ indicates both $\nu_\ell\bar{\nu}_\ell$ and $\ell\ell$.}
    \begin{tabular}{l|c c c c}
    \hline
        \textbf{Final state} &  $\sigma \ [\mathrm{fb}]$& $\epsilon_{presel} [\%]$ & $\epsilon_{tag} \ [\%]$ & $N_{exp}$ \\
    \hline
       $H (\to b \bar{b}) X $   & $490$  & $22.2$  & $32.4$   & $351518$ \\
       \hline
       $H (\to c \bar{c}) X $   & $24.3$ & $22.2$  & $4.49$   & $2422$  \\
       $q\bar{q} \nu_\ell \bar{\nu}_\ell$ & $2674$ & $25.6$  & $5.00$  & $341598$ \\
       $q\bar{q} \ell\ell$ & $4339$ & $1.86$  & $1.31$  & $10533$ \\
       $q\bar{q} \ell \nu_\ell$     & $9763$ & $21.46$   &$0.10$   & $20974$   \\ 
      \hline
    \end{tabular}
    \label{tab:hbb}
\end{table}

\begin{figure}[t]
\centering
\includegraphics[width=0.9\linewidth]{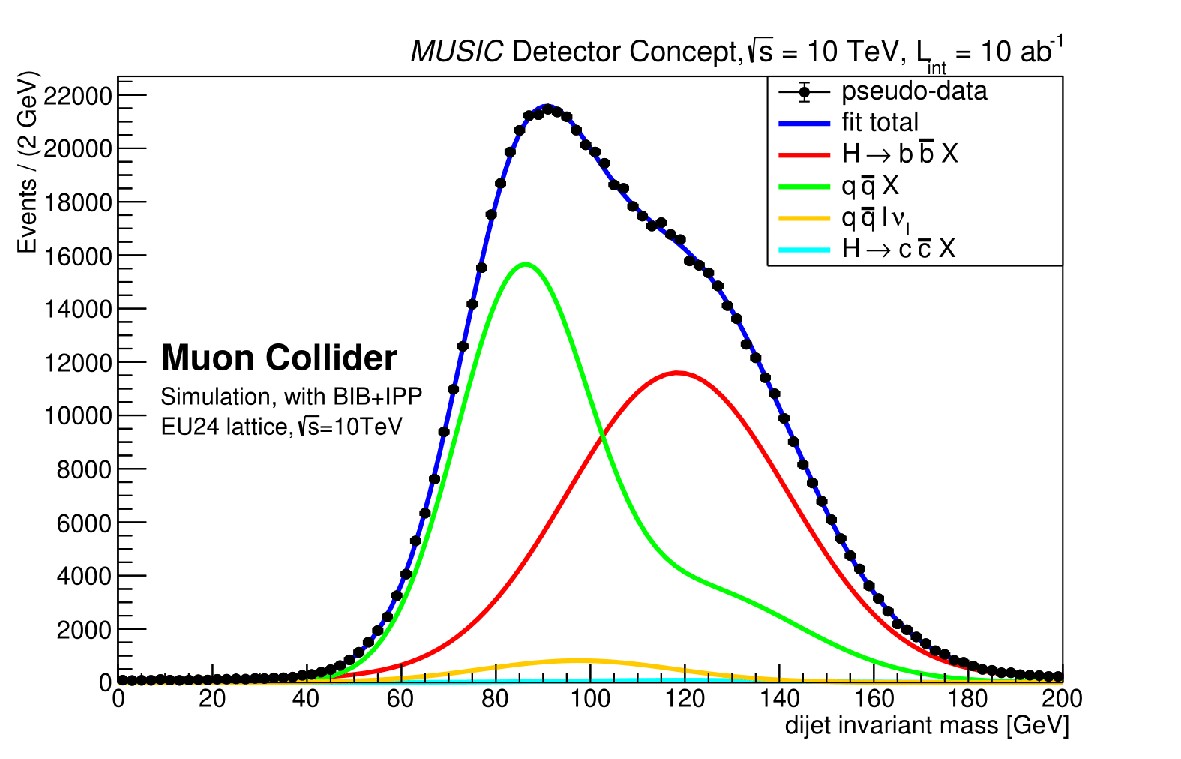}
\caption{Example of the dijet invariant mass fit for one of the pseudo-experiments used to extract the $H \rightarrow b \bar{b}$ yield and uncertainty. 
In the legend, $X$ stands for both $\nu_\ell\bar{\nu}_\ell$ and $\ell\ell$. }
\label{fig:hbb_fit}
\end{figure}

\subsection{$\mathbf{H\rightarrow b \overline b}$}
\label{sect:hbb}

\noindent
The $H\to b \bar{b}$ channel is reconstructed in the dijet final state. The two largest $p_T$ jets in the events are selected, and they are both required to possess a $p_T > 40$ GeV/c. The fake jet removal and \textit{b}-jet identification procedures are described in detail in Ref. \cite{music-det-esppu}. 
The identification efficiencies and mis-tag rates are applied as event weights and as functions of the jet kinematics. The working point of the jet tagging algorithm is chosen to have $b$-jet efficiencies $\sim 55\%$ and $c$-jet and light jet mis-tag rates $\sim 20\%$ and $\sim 1\%$, respectively.
The generation cross sections, the selection efficiencies, the jet identification efficiencies, and the expected number of observed signal and background events are reported in Tab. \ref{tab:hbb}.

\noindent
A fit to the dijet invariant mass distribution is performed to extract the Higgs signal yield (Fig.\ref{fig:hbb_fit}). First, the dijet invariant mass model is determined by considering the signal and background process distributions and their expected yields. The distributions are fitted with double Gaussian probability density functions that are included in the model. Pseudo-experiments are extracted from this model. Then, an unbinned maximum likelihood fit is performed to the pseudo-experiment dijet mass distribution, with the $H \to b \bar{b}$ and $q\bar{q} \nu \bar{\nu}$ yields as free parameters. The other background yields are fixed to the expectation. 
A number of 1000 pseudo-experiments is performed, and a fit to each pseudo-dataset is done. The average uncertainty on the $H \to b \bar{b}$ yield among the pseudo-experiments is taken as the expected uncertainty. 
Assuming negligible uncertainties on the selection efficiency and the integrated luminosity, the relative statistical uncertainty on the $H \to b \bar{b}$ production cross section is directly determined by the statistical sensitivity of the signal yield:
\begin{equation}
    \frac{\Delta (\sigma\cdot BR) (H\to b \bar{b})_{stat}}{(\sigma\cdot BR) (H\to b \bar{b})} = 0.20\%\ .
\end{equation}

\subsection{$\mathbf{H\rightarrow WW^*}$}

\noindent
The $H\to WW^\ast$ channel is reconstructed in the semileptonic mode, where one $W$ boson decays into a muon and a muonic neutrino, while the other decays hadronically into two jets, resulting in the final state: $H\to WW^\ast \to q\bar{q}\mu \nu_\mu$.

\noindent
Jets with $20 < p_T < 2000$ GeV/c and polar angle direction $10^\circ < \theta_\mu < 170^\circ$ are selected. If more than two jets meet these requirements in an event, the jet pair with invariant mass closest to the nominal $W$ boson mass is selected. Muons are required to have $10 \text{ GeV} < p_T^\mu < 1000$ GeV and $10^\circ < \theta_\mu < 170^\circ$. The muon reconstruction procedure is described in detail in Ref. \cite{music-det-esppu}. If more than one identified muon satisfies these requirements, the most isolated muon is selected. Muon isolation is defined relative to the closest jet in the event as $I_\mu = p_T^\mu/p_T^{jet}$ if $\Delta R = \sqrt{\Delta\phi^2 + \Delta\eta^2} < 0.5$, or as $I_\mu = 1$ otherwise (here $\eta$ represents the pseudorapidity and $\phi$ the angle in the plane transverse to the beam).

\begin{table}[t]
    \centering
    \caption{Generation cross-section, preselection efficiency, and expected event yields for signal and background processes in the $H \to WW^{*}$ analysis with $\mu^+\mu^-$ collisions at 10 TeV and $\mathcal{L}=10$ ab$^{-1}$. The symbol $X$ denotes both $\nu_\ell\bar{\nu}_\ell$ and $\ell\ell$.}
    \begin{tabular}{l|c c c}
    \hline
    \textbf{Final state} & $\sigma$ [fb] & $\epsilon_{\textit{presel}}$ [\%] & $N_{\textit{exp}}$ \\
    \hline
    $H (\to W W^* \to q\bar{q} \mu \nu_\mu) X$ &    26.3 & 47.3 &   137493 \\
    \hline
    $ H \nu_\ell\bar{\nu}_\ell$                   &   820   & 12.2 &  1000906 \\
    $H \ell\ell$                               &    84.8 & 12.5 &   106226 \\
    $q\bar{q} \ell\nu_\ell $                  &  9763   & 11.4 & 11110294 \\
    $q\bar{q} \nu_\ell\bar{\nu}_\ell$         &  2674   & 10.2 &  2731663 \\ 
    $q\bar{q} \ell\ell$                       &  4339   &  1.8 &   772342 \\
    \hline
    \end{tabular}
    \label{tab:h2ww}
\end{table}

\noindent
Two Boosted Decision Trees (BDTs) are used to discriminate the signal from the background including a Higgs boson ($HX$), and from non-resonant backgrounds ($q\bar{q}X$), respectively.

\noindent
The statistical sensitivity of the signal yield is evaluated using a toy Monte Carlo study, which fits the two-dimensional distribution of the $HX$ and $q\bar{q}X$ BDT outputs for signal and backgrounds.
An extended binned maximum-likelihood model is defined with three components: the signal $H\to WW^\ast$, $HX$ backgrounds, and $q\bar{q}X$ backgrounds. 
The sensitivity to the signal event count is determined from the mean and width of the distribution of the fitted signal yields in a sample of 10000 pseudo-experiments.
Assuming that the uncertainties on the selection efficiency and the integrated luminosity are negligible, the relative statistical uncertainty on the $H \to WW^*$ production cross section is directly determined by the statistical sensitivity of the signal yield: 
\begin{equation}
    \frac{\Delta (\sigma\cdot BR) (H\to WW^*)_{stat}}{(\sigma\cdot BR) (H\to WW^*)} = 0.41\%\ .
\end{equation}

\begin{figure}[t]
    \centering
    \includegraphics[width=0.9\linewidth]{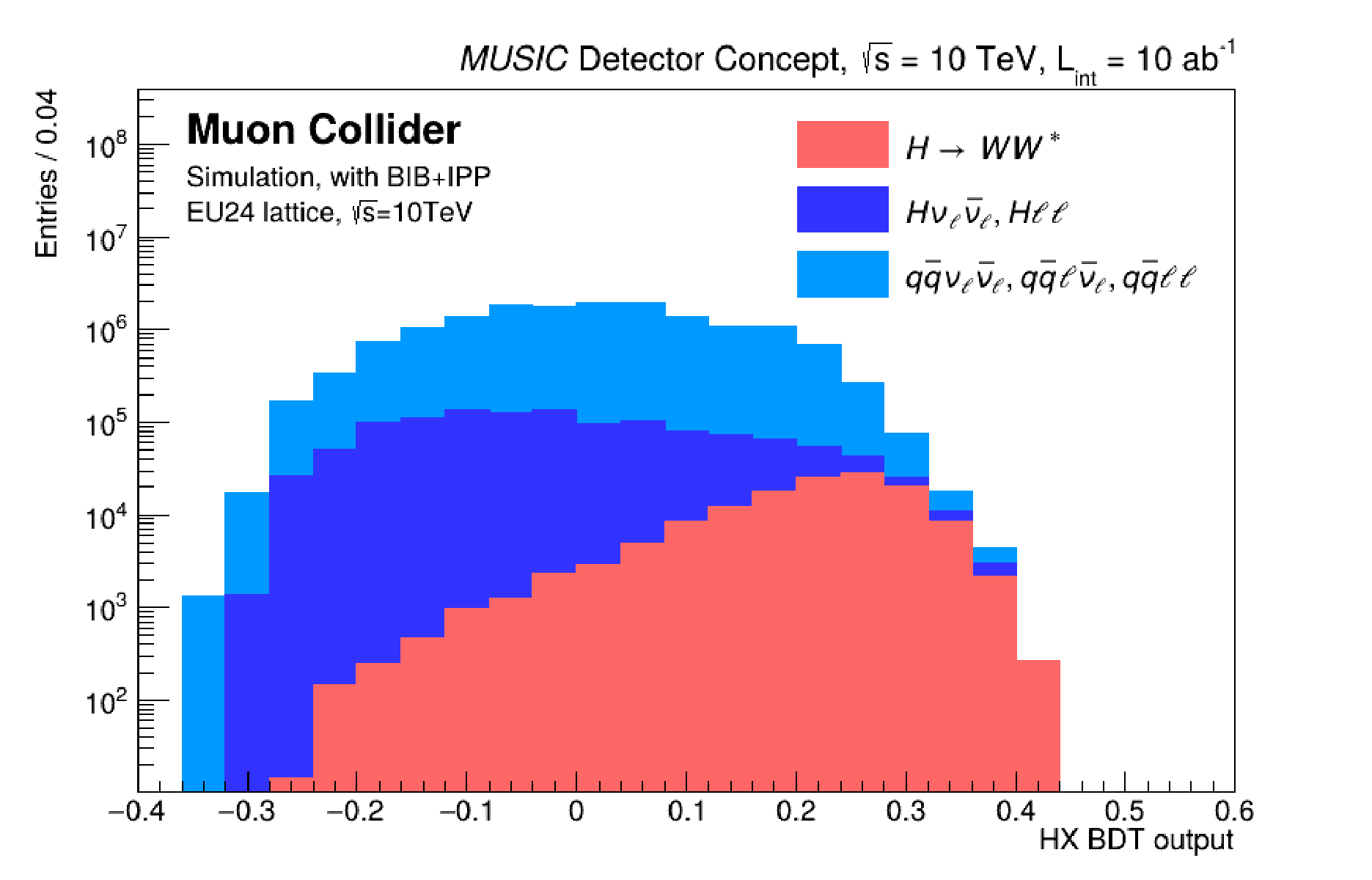}
    \includegraphics[width=0.9\linewidth]{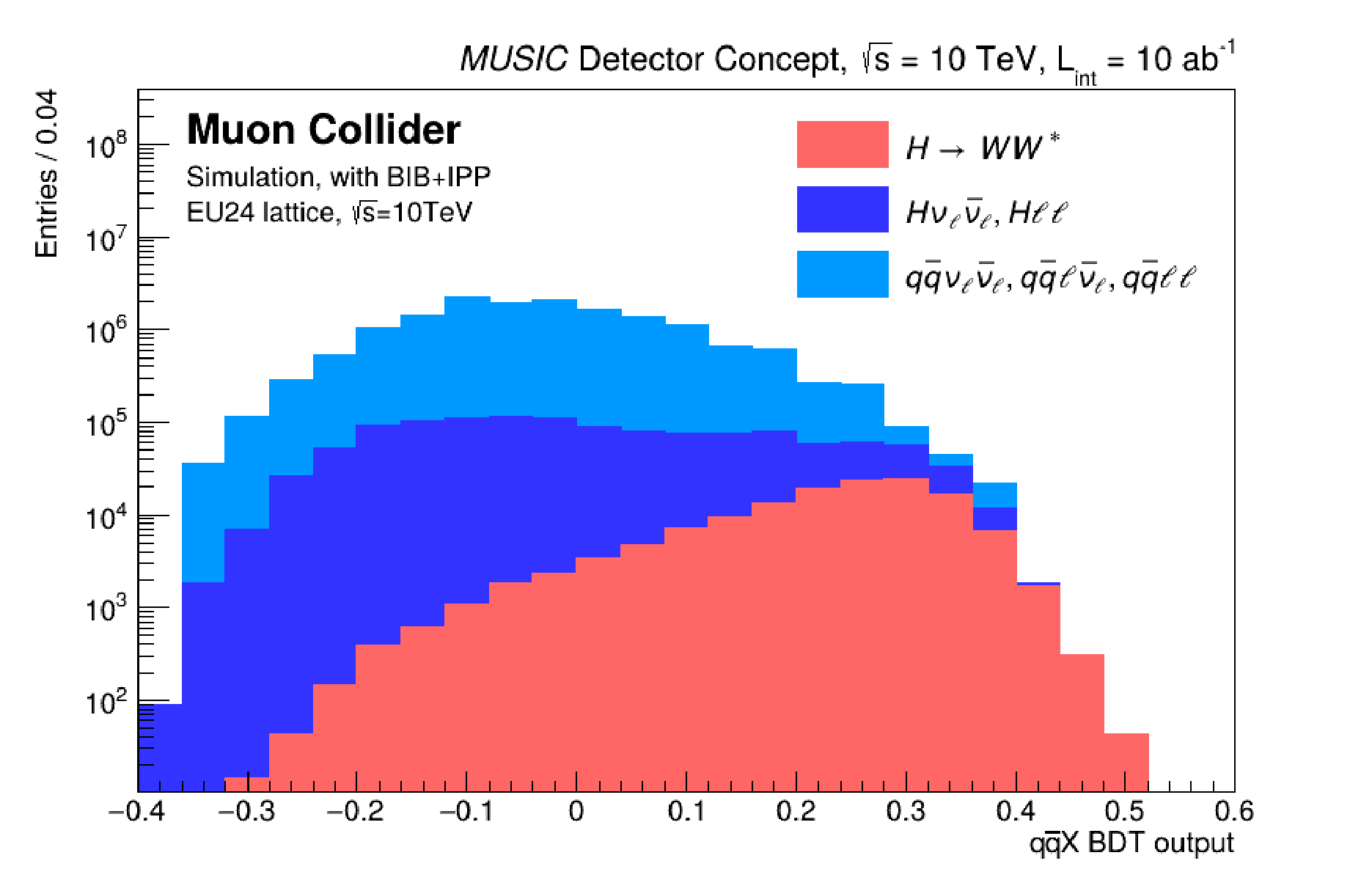}
    \caption{Stacked distributions of the BDT scores to separate the $H\to WW^*$ signal from the backgrounds with a Higgs boson (up) and the non-resonant backgrounds (down).
    \label{fig:h2ww}}
\end{figure}

\subsection{$\mathbf{HH\to b \overline bb \overline b}$}\label{sect:hhbbbb}

\noindent
The $HH\to b \overline bb \overline b$ channel is reconstructed in the four-jet final state. The event pre-selection requires the presence of at least four reconstructed jets, all having $p_T > 20$ GeV/c. The \textit{b}-jet identification efficiencies and mis-tag rates are applied as event weights, as in Section \ref{sect:hbb}.  To reconstruct the $HH$ events, all possible two-jet combinations are formed, in which at least one jet is identified as a $b$-jet.
The two Higgs boson candidates are then built from the two jet pairs whose invariant masses $m_{12}$ and $m_{34}$ minimize the figure of merit:
\begin{equation}
F=\sqrt{(m_{12}-m_{H})^2+(m_{34}-m_{H})^2}\ ,
\label{F_figure_merit}
\end{equation}
where $m_H$ is the nominal Higgs boson mass. The main physics background contribution comes from processes with four heavy-quark jets in the final state, $\mu^+ \mu^- \to q_h \bar{q}_h q_h \bar{q}_h (\nu_\ell\bar{\nu}_\ell, \ell \ell)$  ($q_h=b$ or $c$). Another important source of background is the process $\mu^+ \mu^- \to H  q_h \bar{q}_h (\nu_\ell\bar{\nu}_\ell, \ell \ell)\to b \bar{b}  q_h \bar{q}_h (\nu_\ell\bar{\nu}_\ell, \ell \ell)$ where 
the $q_h \bar{q}_h$ pair is non-resonant. A summary table presenting the considered signal and background processes, along with their cross sections and expected event yields, is provided in Tab.~\ref{tab:HH}, where the efficiency $\epsilon$ takes into account both the preselection and tagging efficiencies.

\noindent
A Multilayer Perceptron (MLP) is used to discriminate the signal from double-Higgs production from the background processes. 

\begin{figure}[t]
    \centering
    \includegraphics[width=0.9\linewidth]{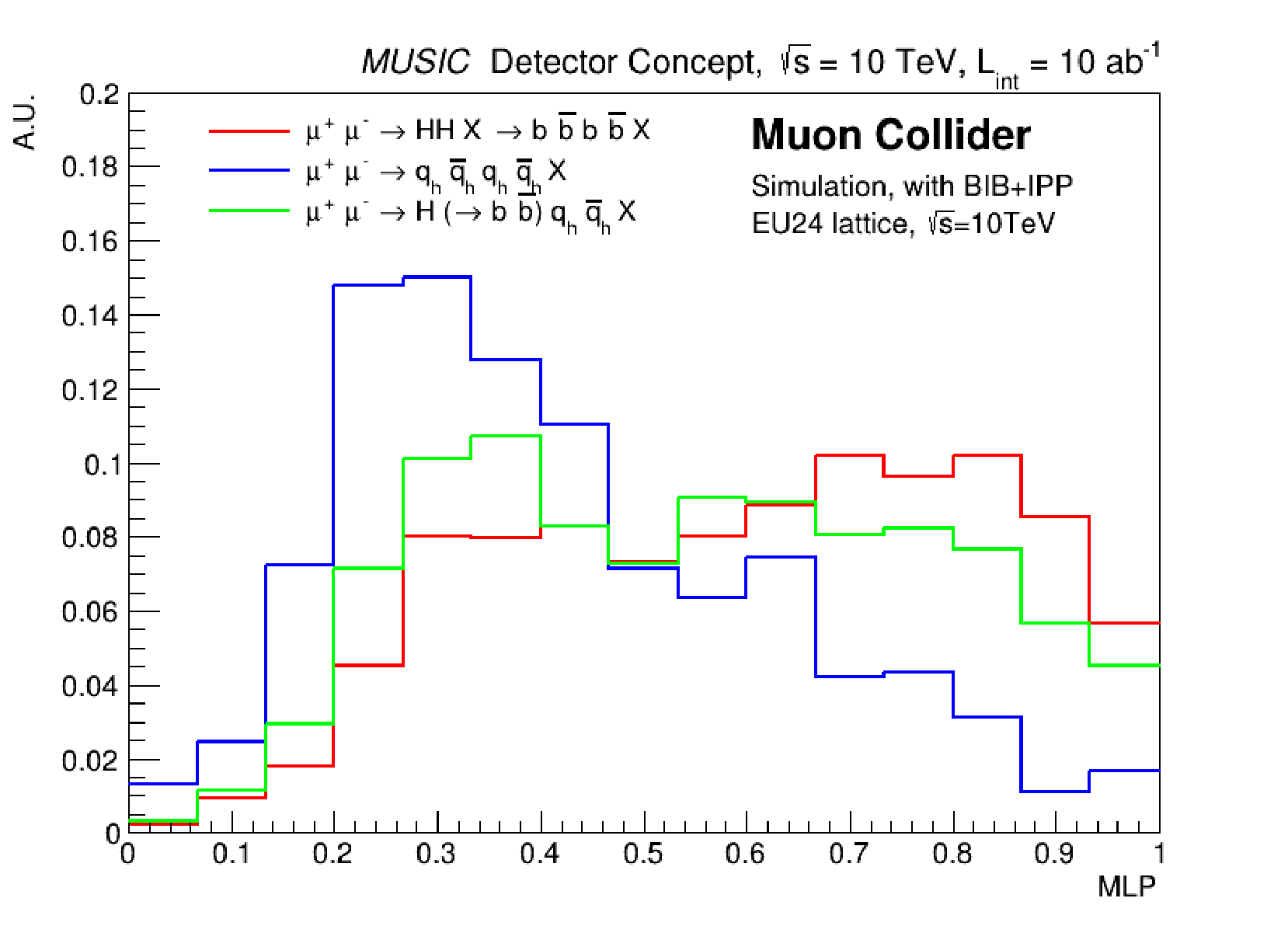}
    \caption{Distributions of the MLP output for the $H\!H$ signal and the main background contributions. The distributions are normalized to unit area.
    \label{fig:MLP_HH}}
\end{figure}

\noindent
The MLP output distributions for the $H\!H$ signal and background samples are shown in Fig.~\ref{fig:MLP_HH}. 
Using the MLP output distributions for $HH$ signal and background samples, pseudo-datasets have been generated according to the expected number of events, and the procedure already presented for $H\to b \bar b$ and $H\to WW^*$ is applied. The yield of the $HH$ signal is extracted by fitting the MLP distribution. Assuming that the uncertainties on the selection efficiency and the integrated luminosity are negligible, the relative statistical uncertainty on the $H H\to b \overline bb \overline b$ production cross section is directly determined by the statistical sensitivity of the signal yield:
\begin{equation}
    \frac{\Delta (\sigma\cdot BR)  (H\!H \rightarrow b\bar{b} b\bar{b})_{stat}}{ (\sigma\cdot BR) (H\!H \rightarrow b\bar{b} b\bar{b})} = 4.2\%\ .
\end{equation}

\begin{table}[t]
    \caption{Generation cross-section, total selection efficiency, and expected events for signal and background processes in the $H\!H \rightarrow b\bar{b} b\bar{b}$ analysis with $\mu^+\mu^-$ collisions at 10 TeV and $\mathcal{L}=10$ ab$^{-1}$. $X$ indicates both $\nu_\ell\bar{\nu}_\ell$ and $\ell\ell$.}
    \centering
    \begin{tabular}{l|c c c c}
    \hline
    \hline
         \textbf{Final state} & $\sigma \ [\mathrm{fb}]$ &  $\epsilon [\%]$ &  $N_{exp}$ \\
    \hline
       $H\!H (\to b \bar{b} b \bar{b}) X$         & $1.14$ & $18.47$ & $2100$ \\
       \hline
       $H (\to b \bar{b}) q_{h}  \bar{q}_{h} X$  & $7.27$ & $15.56 $ & $11307$ \\
       $q_{h}  \bar{q}_{h} q_{h}  \bar{q}_{h} X$ & $10.89$ & $8.99$ & $9787$ \\
      \hline
      \hline
    \end{tabular}
    \label{tab:HH}
\end{table}

\subsection{\textbf{Higgs trilinear}}

The Higgs trilinear coupling enters the double-Higgs production cross-section due to the contribution from the $HHH$ vertex. A direct measurement of the trilinear coupling modifier $\kappa_3$ requires the disentanglement of the contribution of the $HHH$ vertex from the other double-Higgs production modes. This study is performed considering the double-Higgs decay mode $HH \to b\bar{b} b\bar{b}$.

\begin{figure}[t]
    \centering
    \includegraphics[width=0.9\linewidth]{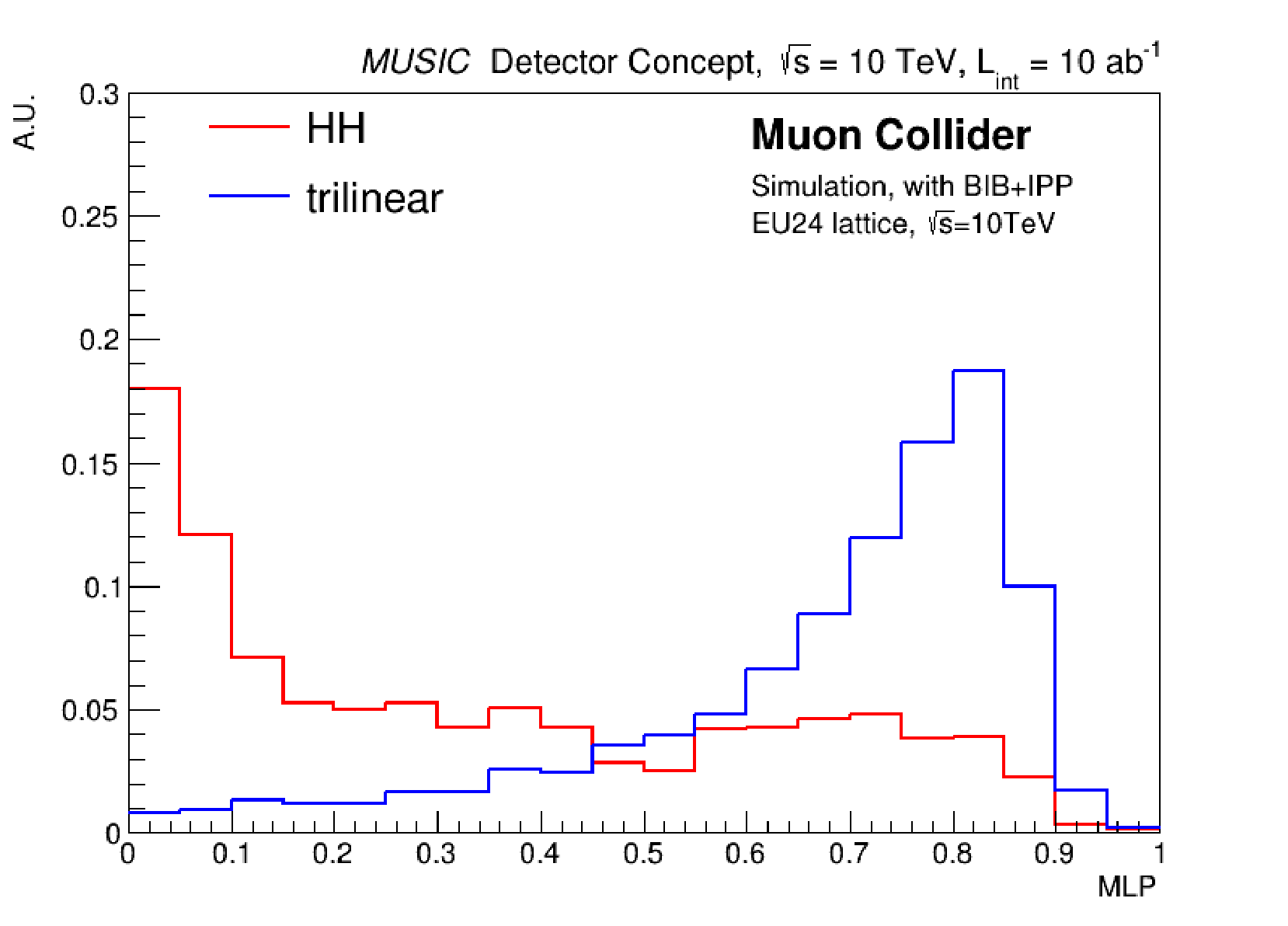}
    \caption{output distributions for the MLP applied to distinguish the $H\!H$ events sensitive to $\lambda_3$ from the other $H\!H$ contributions. \label{fig:mlp2}}
\end{figure}

\begin{figure}[t]
    \centering
    \includegraphics[width=0.9\linewidth]{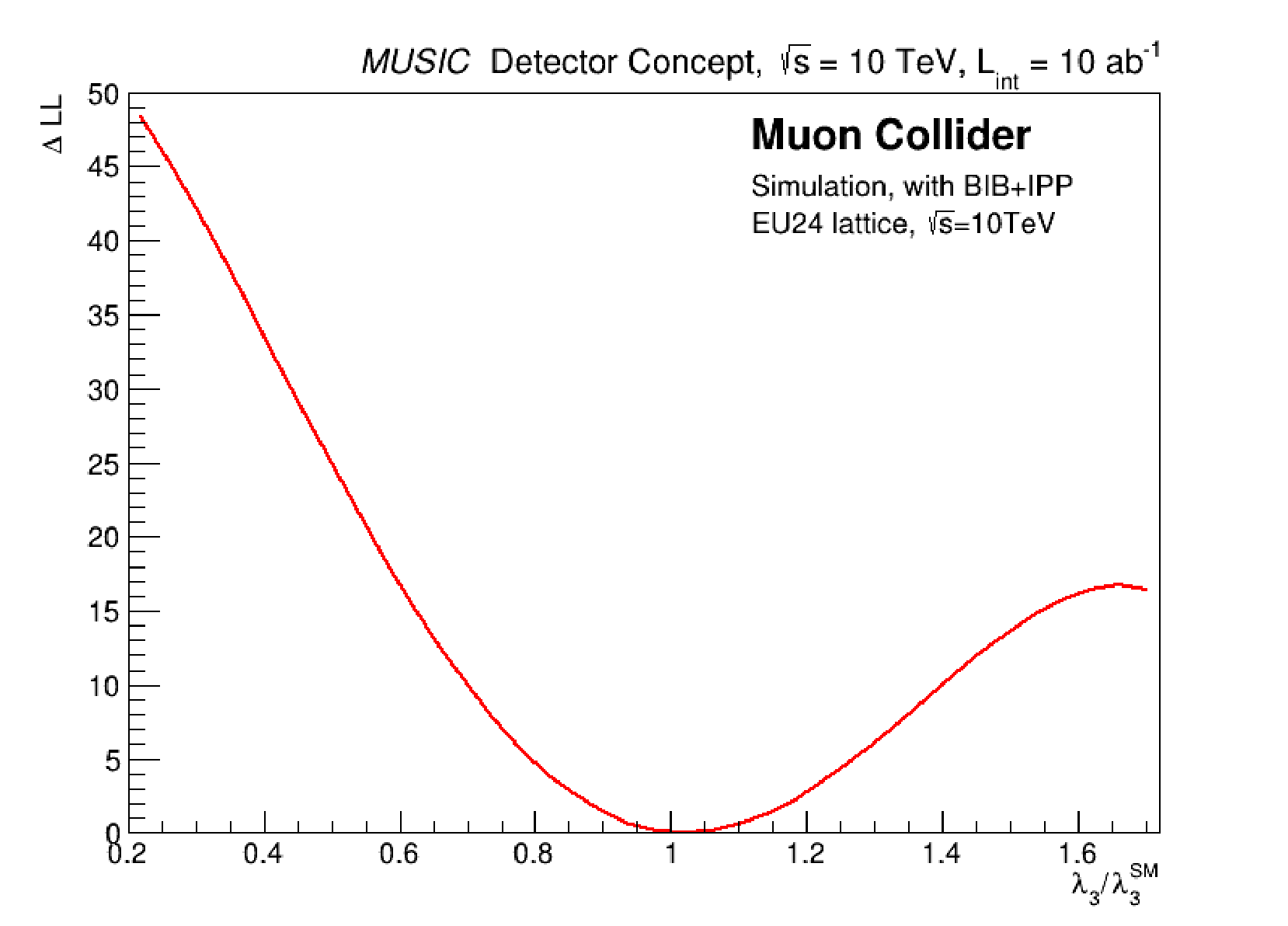}
    \caption{In the left panel,  On the right, $\Delta LL$ as a function of the $\kappa_{\lambda_3}$ hypothesis.
    \label{fig:DLL}}
\end{figure}

\noindent
The procedure to evaluate the sensitivity to $\kappa_3$ is described:
\begin{itemize}
    \item[-] Eleven samples of double-Higgs events are generated with WHIZARD for values of $\kappa_3$ ranging from 0.2 to 1.8, and then simulated and reconstructed, with the $\kappa_3=1$ case representing the SM signal.
    \item[-] Two independent MLPs are exploited:
    \begin{enumerate}
        \item The first MLP discriminates between double-Higgs events and the physics background (the same as Sect. \ref{sect:hhbbbb}).
        \item The second MLP discriminates between the double-Higgs events containing the $HHH$ vertex and the other double-Higgs production modes (Fig.\ref{fig:mlp2}).
    \end{enumerate}

    \item[-] For each $\kappa_3$ hypothesis, 2-dimensional templates for the signal and background components are built using the 2D histograms of the outputs of the two MLPs, normalized to the signal and background yields.

    \item[-] Pseudo-datasets are generated with the total 2D template for the $\kappa_3=1$ hypothesis. For each pseudo-experiment, the likelihood difference $\Delta LL = -\Delta \mathrm{log} (L)$ is calculated as a function of $\kappa_3$  by comparing the pseudo-data distribution to the $\kappa_3$ templates.
    
    \item[-] The log-likelihood profile is fitted with a fourth-degree polynomial function. The 68\% C.L. uncertainty on $\kappa_3$ is determined as the interval around $\kappa_3=1$ where the fitted polynomial has a value less than 0.5. The profile is shown in Fig.\ref{fig:DLL}.
\end{itemize}

\noindent
The likelihood scan results in a confidence interval for the trilinear Higgs self-coupling modifier of $0.96<\kappa_3<1.06$ at 68\% C.L.


\section{Conclusions and outlook}
\noindent
This contribution highlights the potential of a $\sqrt s = 10$ TeV Muon Collider in expanding the present understanding of the Higgs sector. The attainable sensitivities on a representative set of Higgs boson observables are evaluated, using the detailed simulation of the MUSIC detector concept and including the main machine-induced backgrounds.

\noindent
The projected statistical uncertainties on the cross sections times branching ratio ($\sigma\cdot BR$) of the Higgs boson, $0.20\%$ for $H \to b\bar{b}$, $0.41\%$ for $H \to WW^*$ and $4.2\%$ for $H\!H \to b\bar{b}b\bar{b}$, highlight the capability of a Muon Collider to probe the Higgs couplings to fermions and bosons with high precision. The trilinear Higgs coupling modifier is constrained to $0.96 < \kappa_3 < 1.06$ at 68\% C.L., providing a stringent test of the Higgs potential and possible deviations from the Standard Model. The results assume an integrated luminosity of $\mathcal{L} = 10$ ab$^{-1}$, which is expected to be collected by each of the two experiments in five years of operation.

\noindent
Future studies aim to include a larger set of Higgs decay processes and to evaluate the sensitivity to the trilinear coupling modifier in other significant channels, \textit{e.g.} $HH\to b\bar b \, WW^*$, to complete the picture on the potential of a multi-TeV Muon Collider in understanding the Higgs sector.

\bibliography{sn-bibliography}

\end{document}